\pdfoutput=1
\documentclass[reprint,amsmath,amssymb,aps,prl,twocolumn,tightenlines,superscriptaddress]{revtex4-2}
\usepackage{graphicx}
\usepackage{units}
\usepackage{bm}
\usepackage{amsfonts}
\usepackage{amsmath}
\usepackage{amssymb}
\usepackage{makecell}
\usepackage[table]{xcolor}
\usepackage{gensymb}
\usepackage[utf8]{inputenc}
\newcommand{\pic}[2][1.0]{\includegraphics[width=#1\columnwidth]{#2}}

\newcommand{\nuc}[2]{\hbox{$^{#1}$#2}}

\begin{document}

%\preprint{V. 0.5}

\title{Dissipative reactions with intermediate-energy beams -- a novel approach to populate complex-structure states in rare isotopes}



\author{A.\ Gade}
   \affiliation{Facility for Rare Isotope Beams,
      Michigan State University, East Lansing, Michigan 48824, USA}
   \affiliation{Department of Physics and Astronomy,
      Michigan State University, East Lansing, Michigan 48824, USA}
\author{B.\ A.\ Brown}
    \affiliation{Facility for Rare Isotope Beams,
      Michigan State University, East Lansing, Michigan 48824, USA}
    \affiliation{Department of Physics and Astronomy,
      Michigan State University, East Lansing, Michigan 48824, USA}
\author{D.\ Weisshaar}
    \affiliation{Facility for Rare Isotope Beams,
      Michigan State University, East Lansing, Michigan 48824, USA}
\author{D.\ Bazin}
    \affiliation{Facility for Rare Isotope Beams,
      Michigan State University, East Lansing, Michigan 48824, USA}
      \affiliation{Department of Physics and Astronomy,
      Michigan State University, East Lansing, Michigan 48824, USA}
\author{K.\ W.\ Brown}
    \affiliation{Facility for Rare Isotope Beams,
      Michigan State University, East Lansing, Michigan 48824, USA}
      \affiliation{Department of Chemistry,
      Michigan State University, East Lansing, Michigan 48824, USA}
\author{R.\ J.\ Charity}
    \affiliation{Department of Chemistry, Washington
    University, St. Louis, Missouri 63130, USA}
\author{P.\ Farris}
    \affiliation{Facility for Rare Isotope Beams,
      Michigan State University, East Lansing, Michigan 48824, USA}
    \affiliation{Department of Physics and Astronomy,
      Michigan State University, East Lansing, Michigan 48824, USA}
\author{A.\ M.\ Hill}
    \affiliation{Facility for Rare Isotope Beams,
      Michigan State University, East Lansing, Michigan 48824, USA}
    \affiliation{Department of Physics and Astronomy,
      Michigan State University, East Lansing, Michigan 48824, USA}
\author{J.\ Li}
    \affiliation{Facility for Rare Isotope Beams,
      Michigan State University, East Lansing, Michigan 48824, USA}
\author{B.\ Longfellow}
 \altaffiliation{Present address: Lawrence Livermore National Laboratory, Livermore, California 94550, USA}
    \affiliation{Facility for Rare Isotope Beams,
      Michigan State University, East Lansing, Michigan 48824, USA}
    \affiliation{Department of Physics and Astronomy,
      Michigan State University, East Lansing, Michigan 48824, USA}
\author{D.\ Rhodes}
 \altaffiliation{Present address: TRIUMF, 4004 Wesbrook Mall, Vancouver, BC V6T 2A3, Canada}
    \affiliation{Facility for Rare Isotope Beams,
      Michigan State University, East Lansing, Michigan 48824, USA}
    \affiliation{Department of Physics and Astronomy,
      Michigan State University, East Lansing, Michigan 48824, USA}
\author{W.\ Reviol}
	\affiliation{Physics Division, Argonne National Laboratory, Argonne, Illinois 60439, USA}
\author{J.\ A.\ Tostevin}
	\affiliation{Department of Physics, Faculty of Engineering and Physical Sciences, University of Surrey, Guildford, Surrey GU2 7XH, United Kingdom}


\date{\today}

\begin{abstract}
A novel pathway for the formation of multi-particle-multi-hole ($np-mh$) excited states in rare isotopes is reported from highly energy- and momentum-dissipative inelastic-scattering events measured in reactions of an intermediate-energy beam of \nuc{38}{Ca} on a Be target. The negative-parity, complex-structure final states in \nuc{38}{Ca} were observed following the in-beam $\gamma$-ray spectroscopy of events in the \nuc{9}{Be}($\nuc{38}{Ca},\nuc{38}{Ca}+\gamma$)X reaction in which the scattered projectile lost longitudinal momentum of order $\Delta p_{||}=700$~MeV/c. The characteristics of the observed final states are discussed and found to be consistent with the formation of excited states involving the rearrangement of multiple nucleons in a single, highly-energetic projectile-target collision. Unlike the far-less dissipative, surface-grazing reactions usually exploited for the in-beam $\gamma$-ray spectroscopy of rare isotopes, these more energetic collisions appear to offer a practical pathway to nuclear-structure studies of more complex multi-particle configurations in rare isotopes -- final states conventionally thought to be out of reach with high-luminosity fast-beam-induced reactions.

\end{abstract}

%\keywords{}

\maketitle


Beyond the proof of existence of a rare isotope and the determination of its ground-state 
half-life, the energies of excited states are typically the first observables that become 
accessible in laboratory experiments. For excited bound states, depending on their lifetime, 
prompt or delayed $\gamma$-ray spectroscopy is frequently used to obtain precise excitation 
energies from the measured transition energies~\cite{gad15}. In short-lived rare isotopes, 
excited states can be populated efficiently in (direct) nuclear reactions~\cite{obe16} or 
$\beta$ decay~\cite{rub09}, for example, most often exploiting the unique selectivity 
inherent to each of these different population pathways. The selectivity of one- and 
two-nucleon transfer and knockout reactions, or inelastic scattering~
\cite{obe16,gla98,han03,gad08a,goe10,wim18}, often enhances the population of excited states
at moderate spin associated with the single-particle or collective degree of freedom.
Here, we report the novel, complementary in-beam $\gamma$-ray spectroscopy of higher-spin, 
negative-parity states in \nuc{38}{Ca}, observed to be populated in the  
\nuc{9}{Be}($\nuc{38}{Ca},\nuc{38}{Ca}+\gamma$)X inelastic scattering at high momentum loss. From the peculiar 
final states observed, we argue that these complex-structure, projectile excited states are 
formed by the rearrangement of multiple nucleons in a single, highly-energetic projectile-target 
collision, giving access to multi-particle configurations not expected to be in reach of high-luminosity fast-beam reactions.

The reaction channel analyzed here is populated in the same experiment as reported on in Ref.
\cite{gad20} where the focus was on \nuc{40}{Sc} produced in the $pn$ pickup reaction onto the
\nuc{38}{Ca} projectile. Here, we briefly summarize the experimental scheme below and refer the reader to Ref.~\cite{gad20,gad22} for more details. The \nuc{38}{Ca} rare-isotope
beam was produced by fragmentation of a stable \nuc{40}{Ca} beam, accelerated to 140~MeV/nucleon
by the Coupled Cyclotron Facility at NSCL \cite{nscl}. The momentum width transported to the
experiment was restricted to $\Delta p/p =0.25$\%, resulting in 160,000 \nuc{38}{Ca}/s impinging
upon a 188-mg/cm$^2$-thick \nuc{9}{Be} foil located at the target position of the S800 spectrograph
\cite{s800}. The setting subject of this publication ran for less than 40 hours. The constituents of the incoming beam and the projectile-like reaction products were
identified on an event-by-event basis using the S800 analysis beam line and focal plane with the
standard detector systems \cite{s800FP}.  As the magnetic rigidity of the 
S800 spectrograph was tuned for \nuc{36}{Ca}, only part of the outermost (exponential) low-momentum tail of the reacted \nuc{38}{Ca} distribution was transmitted to the focal plane. Specifically, the S800 momentum acceptance at this setting is $p_0 \pm 330$~MeV/c, with $p_0=11.222$~GeV/c. 

\begin{figure}[h]
  \begin{center}
    \pic{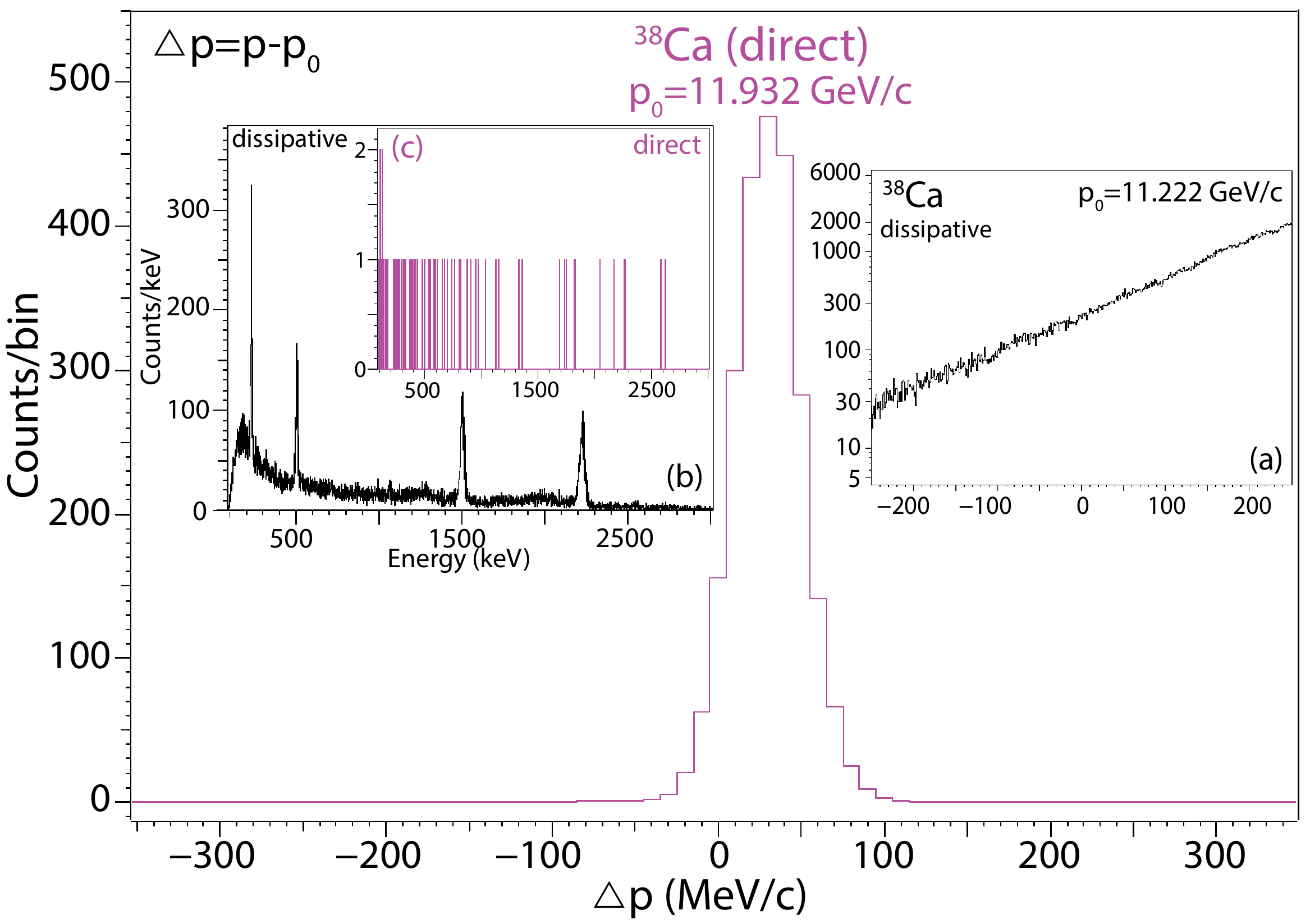}
    \caption{Longitudinal momentum distributions of \nuc{38}{Ca} passing through the target and only suffering energy loss (magenta peak) and, on log scale, for the dissipative setting (inset (a)). Insets (b) and (c) confront the $\gamma$-ray spectra in coincidence with less than 100,500 \nuc{38}{Ca} at high momentum loss (black) and from nearly 179,000 \nuc{38}{Ca} in the direct setting (magenta), highlighting a stark difference in excitation probability.} 
    \label{fig:mom}
  \end{center}
\end{figure}

When compared to the parallel momentum distribution of the unreacted \nuc{38}{Ca} passing through the target, having suffered only in-target energy losses ($p_0=11.932$~GeV/c), the low-momentum, reacted \nuc{38}{Ca} events detected in the reaction setting have undergone an additional longitudinal momentum loss of about 700~MeV/c  (see Fig.~\ref{fig:mom}). That is, approximately 18~MeV/c per nucleon in momentum or 5.4~MeV/nucleon 
in energy. The cross section for finding \nuc{38}{Ca} with such large momentum loss was extracted 
to be $\sigma(p_0 \pm 330~\mathrm{MeV/c})=3.8(4)$~mb, making these inelastic large-momentum-loss events rather rare. 

The mid-target energy of \nuc{38}{Ca} in the \nuc{9}{Be} reaction target was 60.9~MeV/nucleon.
The target was surrounded by GRETINA \cite{pas13,wei17}, an array of 48 36-fold segmented
high-purity germanium crystals assembled into modules of four crystals each, used for prompt
$\gamma$-ray detection to tag the final states of the reaction residues. Signal decomposition
was employed to provide the $\gamma$-ray interaction points. Of these, the location of the
interaction with the largest energy deposition was selected as the first hit entering the
event-by-event Doppler reconstruction of the $\gamma$ rays emitted from the reaction residues
in-flight at about 33\% of the speed of light \cite{wei17}. 

\begin{figure}[h]
\begin{center}
\pic{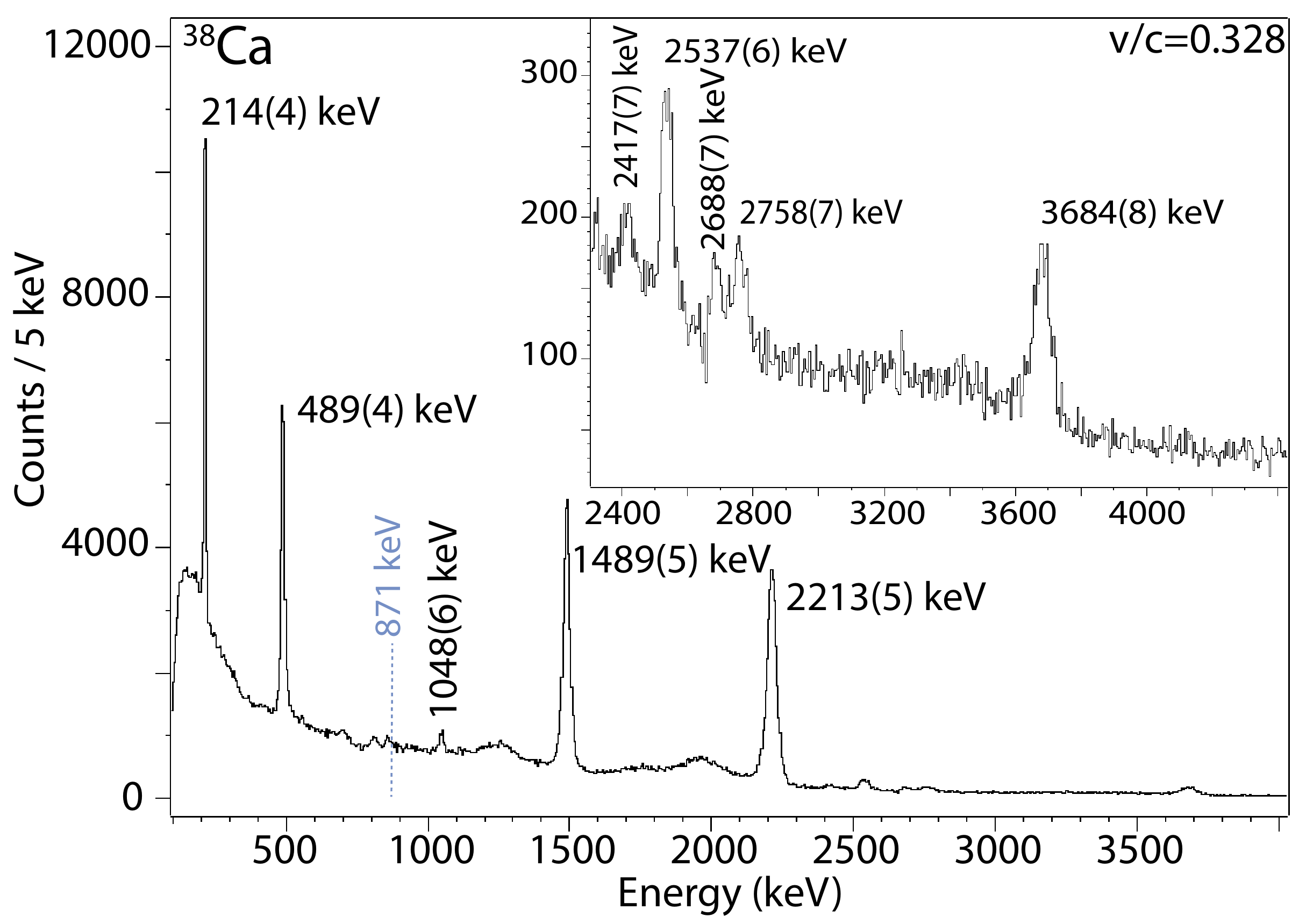}
\caption{Doppler-reconstructed addback $\gamma$-ray spectrum as detected in coincidence with
the scattered \nuc{38}{Ca} nuclei that underwent a large momentum loss. All $\gamma$-ray
transitions are labeled by their energy. The inset magnifies the high-energy region of the
spectrum. }
\label{fig:Ca38_sing}
\end{center}
\end{figure}

The event-by-event Doppler reconstructed $\gamma$-ray spectrum obtained in coincidence with
the \nuc{38}{Ca} reaction residues detected in the S800 focal plane at large momentum loss is
shown in Fig.~\ref{fig:Ca38_sing}. Nearest neighbor addback, as detailed in \cite{wei17}, was
used. Of the seven $\gamma$-ray transitions compiled in \cite{nndc}, those at 2213(5), 1489(5),
489(4) and 3684(8)~keV are observed here, while the transitions at 214(4), 1048(6), 2417(7), 
2537(6), 2688(7), and 2758(7)~keV are reported for the first time in the present work. This letter discusses 
the strongly-populated states. The reader is referred to the companion paper for details 
on some other weakly-populated states \cite{gad22}.

\begin{figure}[h]
\begin{center}
\pic{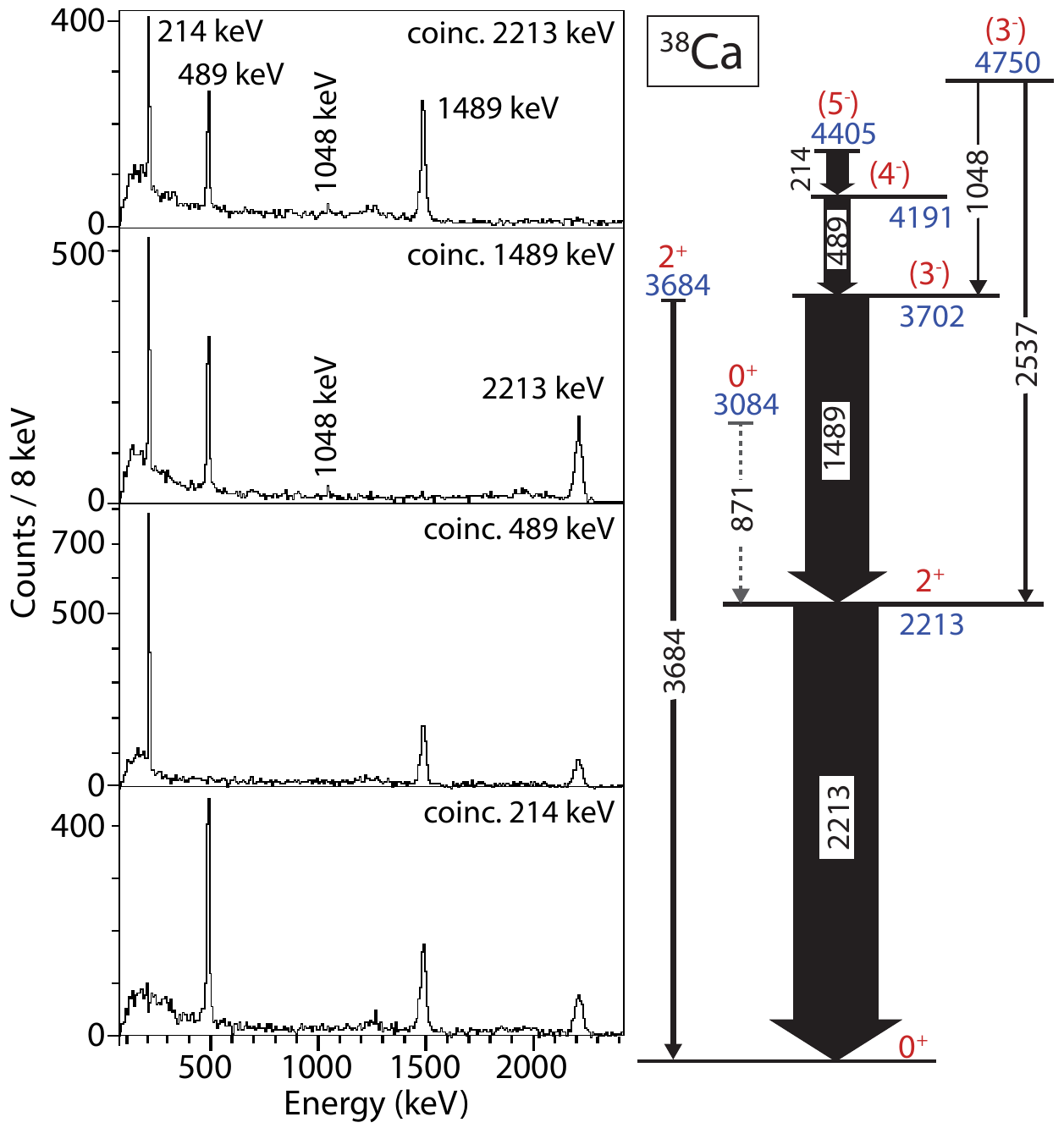}
\caption{Left: Doppler-corrected $\gamma\gamma$ coincidence spectra obtained from cuts on the
labeled prominent transitions in the $\gamma\gamma$ coincidence matrix. Background was subtracted
via a cut of equal width at slightly higher energy. Coincidence relationships are evident in the panels. Right: Resulting level scheme. The width of the arrows is proportional to the
$\gamma$-ray intensity of the corresponding transition. The proton separation energy of $S_p=
4.54727(22)$~MeV \cite{ame20} places the second $3^-$ state above
the proton separation energy. The $0^+_2$ state is shown but was not populated in the present work. }
\label{fig:Ca38_coinc1}
\end{center}
\end{figure}

To construct the level scheme, $\gamma\gamma$ coincidences are used. Figure \ref{fig:Ca38_coinc1}
shows the coincidence analysis of the low-energy part of the \nuc{38}{Ca} spectrum. From 
Fig.~\ref{fig:Ca38_coinc1}, it is clear that the 1489-keV $\gamma$ ray feeds the 2213-keV 
line, the 489-keV transition feeds the level depopulated by the 1489~keV, and the 214-keV 
transition lies on top of the level depopulated by the 489-keV transition. There is evidence for a weak 1048-keV transition being in coincidence with the 2213 and 1489~keV $\gamma$ 
rays.

Figure \ref{fig:Ca38_coinc1} also shows the partial level scheme with the intensities of the
$\gamma$-ray transitions indicated by the arrow widths. These relative 
$\gamma$-ray intensities were deduced from the efficiency-corrected peak areas from the 
spectrum displayed in Fig.~\ref{fig:Ca38_sing}. Remarkably, the fourth strongest $\gamma$ 
ray, at 214~keV, has not been reported previously. The relative intensities 
and a more complete level scheme, including all $\gamma$-ray transitions observed, is 
provided in Ref. \cite{gad22}.

The 1489-keV transition in coincidence with the $2^+_1 \rightarrow 0^+_1$ decay is consistent 
with the previously reported ($3^-$) state at 3702~keV. The 489-keV transition in coincidence 
with the 3702-keV ($3^-$) state suggests a level at 4191~keV, which is consistent with a  
previously reported state at 4194~keV. However, the $J^{\pi}$ assignment proposed in the literature of 
($5^-)$ \cite{nndc} is unlikely as the 489-keV transition in our work is prompt, on 
the level of a few ps or faster as evident from the good resolution and absence of a low-energy 
tail, which -- if of $E2$ character -- would indicate a $B(E2; 5^- \rightarrow 3^-)$ strength 
exceeding the recommended upper limit of 100~W.u. \cite{end93}. From comparison with the mirror 
nucleus, \nuc{38}{Ar}, which has a 4480-keV $4^-$ level with a sole transition of 670~keV 
connecting to the first $3^-$ state, resembling the situation described here, we propose $J^{\pi}=(4^-)$ for the 4191-keV state in \nuc{38}{Ca}. The new 214-keV transition feeding 
the ($4^-$) level establishes a state at 4405~keV which appears to correspond to the 4586-keV
$5^-$ level in the \nuc{38}{Ar} mirror whose far-dominant decay is a 106-keV transition to
the $4^-$. Based on mirror symmetry, a ($5^-$) assignment is proposed here for the
4405-keV level in \nuc{38}{Ca}. This establishes $(5^-) \rightarrow (4^-) \rightarrow (3^-_1)
\rightarrow 2^+_1$  as the most intense cascade seen following the \nuc{38}{Ca} inelastic 
scattering populated at large momentum loss.

%\begin{figure}[h]
%\begin{center}
%\pic[0.60]{fig4.pdf}
%\caption{Proposed level scheme for \nuc{38}{Ca}. The width of the arrows is proportional to the
%$\gamma$-ray intensity of the corresponding transition. The proton separation energy of $S_p=
%4.54727(22)$~MeV \cite{ame20} places the second $3^-$ state above
%the proton separation energy. The $0^+_2$ state is shown but was not populated in the present work.  Spin-parity %assignments in parenthesis are tentative.}
%\label{fig:Ca38_lev}
%\end{center}
%\end{figure}

The next strongest populated level is the $2^+_2$ state at 3684~keV for which only the
transition to the ground state is observed here.  A $0^+$  state at 
4748(5)~keV is claimed in \nuc{38}{Ca} from the $(\nuc{3}{He},n)$ transfer reaction, 
however, with the suspicion of a doublet \cite{nndc}. Due to the transition to the 
$(3^-$) state, a $0^+$ assignment is excluded and the level established here is 
tentatively assigned ($3^-_2$), consistent with the 4877-keV 
$3^-_2$ level in the \nuc{38}{Ar} mirror, which also decays predominantly to the $2^+_1$ 
and $3^-_1$ states \cite{nndc}.

It is interesting to explore which low-lying levels have not been observed in the present
experiment. This is, most prominently, the $0^+_2$ state reported at 3084~keV
which would decay to the first $2^+$ state with a 871-keV transition \cite{nndc}. There is
no evidence for an appreciable presence of that transition in Figs.~\ref{fig:Ca38_sing} and \ref{fig:Ca38_coinc1} (the 871-keV transition would be 13~keV above the background feature originating from neutron-induced background as indicated in Fig.~\ref{fig:Ca38_sing}).

In the following, we discuss the configurations of the states observed. Many properties of 
\nuc{40}{Ca} and the surrounding nuclei can be interpreted relative to a doubly-closed shell 
structure for the ground state of \nuc{40}{Ca} with the $sd$ shell filled and the $pf$ shell 
empty. The first excited state of \nuc{40}{Ca} has $  J^{\pi }=0^{+}  $ and is qualitatively 
associated with a four-particle four-hole (4p-4h) state relative to the \nuc{40}{Ca} 
closed-shell ground state \cite{ger67}. We will use $\Delta$, the number nucleons moved from $sd$ to $fp$ orbitals, to characterize the structure of the states. In this notation, 
the 4p-4h states in \nuc{40}{Ca} have $\Delta=4$. (To remove spurious states, the $\Delta$ 
basis includes all components associated with the $\Delta \hbar \omega$ basis constructed 
in the $0s$-$0p$-$0d1s$-$0f1p$ model space).

\begin{figure}[h]
  \begin{center}
    \pic{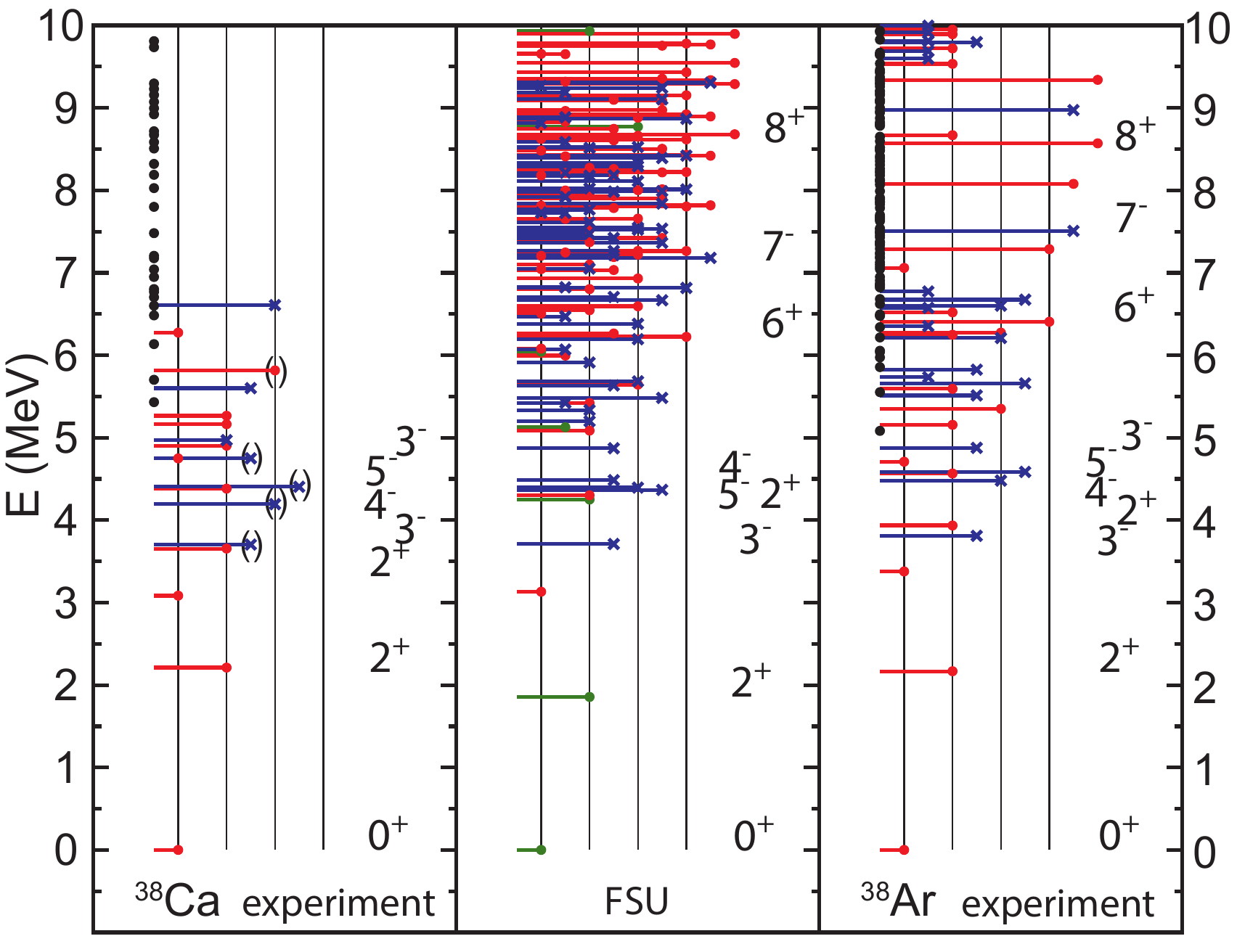}
    \caption{Comparison of the energies of the low-lying states of \nuc{38}{Ca}, with the states observed here labeled, with shell-model calculations using the FSU $s psd fp$ interaction, and states in \nuc{38}{Ar} \cite{nndc}. In these plots, the length of the levels indicates the $J$ value and the color positive parity, $\Delta=2$ (red), negative parity, $\Delta=1$ (blue), and $sd$-shell origin, $\Delta=0$ (green).}
    \label{fig:Ca38_theo}
  \end{center}
\end{figure}

In Ref. \cite{wa90}, a Hamiltonian was developed for these pure $\Delta$ configurations. 
This Hamiltonian served as the starting point for the new Florida State University (FSU) 
Hamiltonian for pure $\Delta$ states \cite{lub19,lub20}. The $A=38$, FSU results are compared 
to experiment in Fig. \ref{fig:Ca38_theo}, the overall agreement with experiment being good. 
The calculated configurations can be divided into those with $\Delta=0$ with positive parity 
(green), those with $\Delta=1$ with negative parity (blue) and those with $\Delta=2$ with 
positive parity (red).

In the present \nuc{38}{Ca} level scheme, the strongest $\gamma$ rays come 
from the $2^+_1$ state, which is predicted to be of $sd$-shell origin, and from states with $\Delta$=1, including the highest $  J^{\pi }=5^{-}  $ level possible 
for this $\Delta$. The $\gamma$-ray decay of the 2$^{ + }_{2}$ state is also observed. In the \nuc{36}{Ar}$(\nuc{3}{He},n)$ reaction in \cite{alf86} this state is found to have a strong $(f_{7/2})^2$ form factor which would come from $\Delta=2$ configurations in the FSU spectrum. However, the 0$^{ + }_{2}$ state, which also has $\Delta=2$, was not populated.

In the following, we propose a view that puts the populated states within the context of 
the observed high-momentum-loss reaction events. From the approximately 200~MeV of energy 
loss in the reaction, and given that the detected \nuc{38}{Ca} are largely within laboratory
scattering angles of 3-4$^{\circ}$, about 150~MeV must be dissipated in the \nuc{9}{Be} 
nuclei, with a total binding of 58~MeV. Thus, there must be disintegration of the target nucleus 
into a number of energetic fragments. The emerging picture is then one of multiple nucleons 
interacting in a single collision with the formation of complex multi-particle multi-hole 
configurations, in contrast to the situation in far-less-dissipative, surface-grazing 
collisions. We exclude scenarios where a \nuc{38}{Ca} projectile undergoes multiple collisions 
within the target as an explanation for the observed cross sections. High-momentum loss events 
creating $mp$-$nh$ excitations in such a scenario would require a sequence of knockout and/or 
pickup processes and such pickup mechanism cross sections are small -- with a typical upper 
limit of 2~mb at these beam energies \cite{gad16}. 

Connecting to the shell-model picture, excitations within the FSU model space are described 
by many-body transition densities. In the simplest scenario, excitation of the $\Delta=1$ 
negative-parity states involve the $\Delta=0$ to $\Delta=1$ one-body transition densities 
(OBTD). The OBTD to those states observed are all large. The $\Delta=2$, 2$^{ + }_{2}$ state 
involves the $\Delta=0$ to $\Delta=2$ two-body transition density (TBTD). The TBTD connecting 
the $\Delta=0$ and $\Delta=2$ 0$^{ + }$ wave functions are the same ones that enter into the 
Hamiltonian matrix for mixing these two states. We expect that the microscopic, two-nucleon 
excitation mechanism should involve an operator similar to that of the two-body mixing 
Hamiltonian (e.g. dominated by pairing). This would explain why excitation of the 0$^{ + }_{2}$ 
is not observed -- the mixed 0$^{ + }_{1}$ and 0$^{ + }_{2}$ eigenfunctions are orthogonal with respect to the two-nucleon
excitation operator. We note that in \nuc{40}{Ca}($p,t$) \cite{kub77} the $0^+_2$ state is only very weakly 
populated compared to the $2^+_2$ state (see Fig. 1 in Ref. \cite{kub77}).

The events at momentum losses of 600-700~MeV/c, studied here, are also reminiscent of 
observations in the work of Podolyak {\it et al.} \cite{pod16}. There, in the two-neutron 
knockout from \nuc{56}{Fe} to \nuc{54}{Fe} at 500~MeV/nucleon, the population of a 10$^+$ 
isomer of complex structure was observed in the low-momentum tail of the parallel momentum 
distribution at about the same absolute momentum loss. The authors attributed this population 
to the excitation of the $\Delta$(1232) resonance at their relativistic beam energies. This 
mechanism is not available to our intermediate-energy beams of tens of MeV/nucleon. One may 
speculate that the population of the complex-structure state  in the two-neutron knockout 
from \nuc{56}{Fe} is rather due to a simultaneous multi-nucleon rearrangement as hypothesized 
here, without evoking quark degrees of freedom and consistent with the reduction of multi-step 
processes at their relativistic energies. For example, population of the $10^+$ state could 
be due to the $\Delta J=6$ excitation of a  $(f_{7/2})^2$ $6^+$ configuration in \nuc{56}{Fe} combined with the removal of two neutrons from the $1p_{3/2}$ and $0f_{7/2}$ orbitals having $\Delta J \geq 4$.

In Ref. \cite{gad22}, from the high-spin spectroscopy of states up to 
$J=15/2$ in \nuc{39}{Ca}, we argue that such simultaneous multi-nucleon rearrangement is also 
at play in intermediate-energy nucleon transfer reactions, such as \nuc{9}{Be}($\nuc{38}{Ca}^*,
\nuc{39}{Ca}+\gamma$)X. Once again, these excitations are seen in events in the tail of 
the longitudinal momentum distribution at a momentum loss of 600-700~MeV/c. 
%Also relevant 
%are the observations of more complex-structure states in the dissipative, low-momentum 
%tails of fast-beam one-nucleon knockout reactions.

In the present work, the specific reaction dynamics at play in the observed large momentum 
loss collisions are unclear and remain a challenge for future, more complete and exclusive 
measurements. Specifically, it would be critical to detect the dissociation of the 
\nuc{9}{Be} target nuclei in the large-momentum-loss events and clarify the kinematics 
of the residues.

While there is much to be discovered about this type of reaction, it is evident that this presents a new opportunity in the fast-beam regime which uniquely complements classic low-energy reactions, such as multi-step Coulomb excitation and multi-nucleon transfer. Fast beams allow for the use thick targets and capitalize on an increase in $\gamma$-ray yield by a factor of about 4300 for the specific example of a 188-mg/cm$^2$ \nuc{9}{Be} target used here vs. a 1-mg/cm$^2$ Pb target often employed for multi-step Coulomb excitation, for example. Also, strong forward focusing enhances the collection efficiency as compared to low-energy reactions that fill a larger phase space. 

Multi-step Coulomb excitation studies with low-energy rare-isotope beams have been performed at beam intensities similar to those used here, but have been limited to a complementary level scheme selectively comprising cascades connected by strong $E2$ transitions, with at most the first $3^-$ state \cite{roc21,goe10}. We illustrate this with the example of the state-of-the art low-energy Coulomb excitation of the neighboring Ca isotope \nuc{42}{Ca} on Pb \cite{had18}. The measurement was performed at 1~pnA stable-beam intensity for 5 days (resulting in more than 110,000 times the number of Ca projectiles on target as in the present measurement) -- excited states up to the $4^+_{1,2}$ states were reported with no evidence for any of the negative-parity cross-shell excitations observed here. 

Multi-nucleon transfer, largely limited to stable beams at pnA beam intensities, is known to populate complex-structure states, however, without efficiently reaching \nuc{38}{Ca} in spite of \nuc{40}{Ca} being an often-used beam (see \cite{cor13} and references within). When low-energy neutron-rich beams become available at near stable-beam intensities, multi-nucleon transfer may become an alternative to access such states in selected neutron-rich nuclei \cite{col20}. While it is interesting to also extend our approach to collective nuclei, it already promises to be a unique method to probe cross-shell excitations near magic numbers, elucidating shell evolution in rare isotopes and exploring the necessary model spaces for a region's description on the quest for a predictive model of nuclei.

In conclusion, the in-beam $\gamma$-ray spectroscopy is reported of higher-spin, complex-structure 
negative-parity states in \nuc{38}{Ca} populated in highly-dissipative processes induced by 
a fast \nuc{38}{Ca} projectile beam reacting with a \nuc{9}{Be} target. This work constitutes the 
first high-resolution $\gamma$-ray spectroscopy of \nuc{38}{Ca} with a modern HPGe 
$\gamma$-ray tracking array. The final states 
observed in the inelastic scattering, \nuc{9}{Be}($\nuc{38}{Ca},\nuc{38}{Ca}+\gamma$)X, at 
large momentum loss are characterized through their particle-hole character relative to the 
\nuc{40}{Ca} closed-shell ground state. Excellent agreement is obtained with shell-model 
calculations using the FSU cross-shell effective interaction. Based on the strongly populated negative-parity states and 
the non-observation of the first excited $0^+_2$ state, we propose a consistent picture 
in which these multi-particle multi-hole states are formed by simultaneous rearrangement 
of multiple nucleons in a single, highly-dissipative collision. These reaction processes, 
seen here in the extreme low-momentum tail of \nuc{38}{Ca}+\nuc{9}{Be} inelastic scattering, 
identify a new pathway to gain access to excited states not usually observed in fast-beam
induced reactions and likely out of reach for low-energy reactions.

This work was supported by the U.S. National Science Foundation (NSF) under Grants No. 
PHY-1565546 and PHY-2110365, by the DOE National Nuclear Security Administration through 
the Nuclear Science and Security Consortium, under Award No. DE-NA0003180, and by the 
U.S. Department of Energy, Office of Science, Office of Nuclear Physics, under Grants 
No. DE-SC0020451 (MSU) and DE-FG02-87ER-40316 (WashU) and under Contract No. DE-AC02-06CH11357 (ANL).  GRETINA was funded by the DOE, Office of Science. Operation
of the array at NSCL was supported by the DOE under Grant No. DE-SC0019034. J.A.T. 
acknowledges support from the Science and Technology Facilities Council (U.K.) 
Grant No. ST/V001108/1.

\end{document}